\documentclass[11pt]{article}

\usepackage{achemso}
\usepackage{graphicx}
\usepackage{soul}

\begin{document}

\title{Hydrophobic and Ionic Interactions in Nano-sized Water
Droplets}
                                                                               
\author{S. Vaitheeswaran$^{1}$ and D. Thirumalai$^{1,2,*}$,\\
$^1$Biophysics Program, Institute for Physical Science and Technology,\\
$^2$Department of Chemistry and Biochemistry,\\
University of Maryland, College Park, MD 20742\\
Email: {\tt thirum@glue.umd.edu}}
                                                                               
\date{\today}

\maketitle

\begin{abstract}
 A number of situations ranging from protein folding in confined spaces, lubrication in tight spaces, and chemical reactions in confined spaces require understanding water-mediated interactions.   As an illustration of the profound effects of confinement on hydrophobic and ionic interactions we investigate the solvation of methane and methane decorated with charges in
spherically confined water droplets.
Free energy profiles for a single methane molecule in droplets, ranging in
diameter ($D$) from 1 to 4 nm, show that the droplet surfaces are strongly favorable
as compared to the interior. From the temperature dependence of the free energy
in $D=3$ nm, we show that this effect is entropically driven.
The potentials of mean force (PMFs) between two methane molecules show
that the solvent separated minimum in the bulk is {\it completely absent} in confined water,
independent of the droplet size since the solute particles are primarily associated with the droplet surface. The tendency of methanes with charges (M$_{q+}$ and
M$_{q-}$ with $q^+ = |q^-| = 0.4e$, where $e$ is the electronic charge) to be pinned
at the surface depends dramatically on the size of the water droplet. When $D=4$ nm,
the ions prefer the interior whereas for $D<4$ nm the ions are localized at the surface,
but with much less tendency than for methanes.
Increasing the ion charge to $e$ makes the surface strongly
unfavorable. Reflecting the charge asymmetry of the water molecule, negative
ions have a stronger preference for the surface compared to positive ions of the same
charge magnitude. With increasing droplet size, the PMFs between M$_{q+}$ and M$_{q-}$
show decreasing influence of the boundary
due to the reduced tendency for surface solvation. We also show that as the solute charge density decreases the 
surface becomes less unfavorable.  The implications of our
results for the folding of proteins in confined spaces are outlined.
\end{abstract}


\section{Introduction}
Hydrophobic interactions are presumed to be the dominant force in the folding of proteins
and the assembly of oligomeric structures \cite{Dill_Biochem:90,Liu_Berne_Nat:05}. The nature of hydrophobic
interactions in proteins has been clarified using experiments which
measure the free energy of transfer
of amino acid side chains from a reference solvent to water
\cite{Tanford_AdvProtChem:70}.  Indeed, understanding the nature of
hydrophobic interactions between small ($<1$ nm) solutes has given a
qualitative perspective on the major forces that render proteins
(marginally) stable \cite{Dill_Biochem:90}.  Interactions between small
hydrophobic
solutes have long served as model systems in describing the balance of forces that stabilize
proteins and peptides. The distance dependent potential of mean force (PMF) between solutes, such as
methane, are quantitatively understood \cite{Pangali_Berne_JCP:79,Pratt_Chandler_JCP:77,
Pratt_AnnRevPhyChem:02,Chandler_Nat:05}. The PMF is a measure of the effect of the solvent
(water) on the mutual interaction between hydrophobic solutes. The PMF between two methanes
has a primary minimum at $\approx 0.38$ nm \cite{Pangali_Berne_JCP:79,Shimizu_JCP:00} that roughly corresponds
to the distance of closest approach in the gas phase. In addition, there is a secondary
solvent-separated minimum at $\approx 0.7$ nm in which the two methane molecules are separated
by a water molecule \cite{Pangali_Berne_JCP:79}. The barrier separating
the two minima represents, perhaps, the desolvation penalty that needs to be paid to squeeze
out water molecules so that a stable hydrophobic core can be created in the protein folding
process \cite{Rank_Baker_ProtSci:97,Cheung_Onuchic_PNAS:02}.

Although the nature of hydrophobic forces between small solutes in bulk
water is understood, relatively little
is known about their interactions in confined
droplets.  The nature of water-mediated interactions between amino acids
in nanopores affects the stability of confined proteins
\cite{Betancourt_Thirumalai_JMB:99,Zhou_Dill_Biochem:01,Klimov_Thirumalai_PNAS:02,Takagi_PNAS:03,
Jewett_Baumketner_Shea_PNAS:04}. Several experiments
have shown that the stability of the folded state, compared to
the situation in the bulk, increases upon confinement \cite{Eggers_Valentine_JMB:01,Ravindra_JACS:04,
Campanini_Bettati_ProtSci:05,Bolis_JMB:04}. In many cases the confinement-induced
stability is adequately explained in terms of entropic destabilization of the unfolded states
\cite{Betancourt_Thirumalai_JMB:99,Zhou_Dill_Biochem:01,Klimov_Thirumalai_PNAS:02}.
However, one can envision scenarios in which alterations in hydrophobic interactions in
confined water can also destabilize the folded state \cite{Cheung_Thirumalai_JMB:06}.
The alteration in the properties of confined water are also important in the context of lubrication
in thin films \cite{Raviv_Sci02} and colloid science. Thus, in  a number of applications the role
highly confined water plays in affecting solvation of hydrophobic and charged species
is important.

Motivated by the above considerations, we have calculated PMFs between two
methane molecules in confined spherical water droplets ranging in diameter from $1-4$ nm.
Interestingly, we find that for all values of the confining diameter $D$, {\it the
solvent-separated minimum is completely destabilized} because  methane molecules are pinned to the
boundary of the sphere. In order to describe the competition between charged residues, which
would prefer to be fully solvated, and hydrophobic interactions, we have
also calculated PMFs between M$_{q+}$ and M$_{q-}$. Here, one of the methane
molecules has a positive
charge $q^+$ while the other carries a negative charge. The extent of destabilization of the
secondary minimum in the confined space depends critically on the magnitude of the charges,
$q^+ = |q^-|$ and the ionic radius.
The possible implications of our results for protein stability in confined geometries are
briefly outlined.

\section{Methods}
{\it Models}: We use Metropolis Monte Carlo \cite{MRRTT'53} simulations to study the energetics
of aqueous solvation of methane molecules and the model ions M$_{q+}$ and M$_{q-}$ in
confined spherical water droplets. The droplet diameter ($D$)   
ranges from
1.0 to 4.0 nm, and are bounded by hard walls. The potential energy at the wall is large
enough ($10^{12}$ kJ/mol) that it confines the system to the desired volume. To simulate
confinement effects we do not use periodic boundaries.  
Electrostatic and Lennard-Jones
interactions are evaluated without a cutoff.  We use
the TIP3P model for water \cite{JCMIK'83} and a unified atom representation for
methane \cite{Kalra_JPC:04}. The Lennard-Jones (LJ) parameters for these interactions are
listed in Table~\ref{tab:LJ_parms}. Model ions are created by adding charges of
magnitude $q^+ = |q^-| = 0.4e$ or $e$ to the methane spheres, where $e$ is the
electronic charge \cite{Wallqvist_Thirumalai_JACS:98}. In order to assess the importance of charge
density in affecting solvation  we also study, for the 3 nm droplet,
ions M*$_{q+}$ and M*$_{q-}$ with $q^+ = |q^-| = e$ and twice the volume of the
methane sphere. The parameters for their LJ interactions with water oxygens are
given in Table~\ref{tab:LJ_parms}. Simulations are performed at constant number
of molecules $N$, total volume $V$ and absolute temperature $T$, and thus sample
the canonical ensemble. The simulations are carried out at 298 K and a water
density of 997~kg/m$^3$.  We also simulated the $D=3$ nm droplet at
a temperature of 328 K with  one methane, at the same water density.

The number of water molecules $N_w$, in each droplet, is calculated by assuming
that the effective volume available to the water is
$$ V_{eff} = V - N_mV_m$$ where $N_m$ is the number of solute molecules in the
droplet and $V_m = (4/3)\pi \sigma_{MO}^3$, is the excluded volume due to each
solute. With the experimental water density of 997~kg/m$^3$ at 298 K and
1 atm pressure, we get the number of water molecules $N_w$ listed in Table~
\ref{tab:N_w}.

{\it Simulation details}:
We performed canonical ensemble Monte Carlo simulations with $N_m$
ranging from 0 to 2 and the appropriate number of waters (Table~\ref{tab:N_w}).
Each system was equilibrated for at least 10$^7$ Monte Carlo steps.
Trial moves consisted of a random translation for both solute and solvent molecules
and an additional random rotation for solvent molecules only. These were accepted
according to the Metropolis criterion \cite{MRRTT'53}.

The free energy of a single methane is calculated as a function of its radial
distance from the center of the droplet using umbrella sampling.
The PMFs between two methane molecules and M$_{q+}$ and M$_{q-}$
confined in these droplets are evaluated using the same technique.
We calculated the PMFs using the weighted histogram analysis method (WHAM)
\cite{Kumar_JCC:92} with code from Crouzy et~al.~\cite{Crouzy_JCC:99}.

We used harmonic biasing potentials, $U_{bias}=k(r-r_0)^2$, where $r_0$ is
the center of each window and the spring constant $k$ is chosen to be 5 kcal/(mol-\r{A}$^2$)
\cite{Roux_CPC:95}. Window centers are 0.1 nm apart in all calculations.
Data sets consist of $2.10^4-10^5$ data points in every window.
Each such data set is added to the previous ones and the WHAM equations
\cite{Kumar_JCC:92} are iterated to a tolerance of $10^{-4}-10^{-5}$
\cite{Crouzy_JCC:99}. The calculations are considered to have converged
if the addition of the last data set, with at least $5.10^4$ data points in each
window, does not change the final PMF significantly. Typically, between
2.10$^7$ and 5.10$^7$ Monte Carlo steps are required to achieve convergence.

We adapted a standard method to calculate the radial distribution functions
$g(r)$, for the confined water. An inner sphere of radius
$r_i=0.5$ nm (0.2 nm for the smallest droplets) was defined in each case and
the $g(r)$'s were averaged for all the waters inside this inner sphere.
The normalization is therefore accurate up to a distance of $R-r_i$, where
$R=0.5D$ is the radius of the droplet. These functions go to 0 at a
distance of $R+r_i$, unlike the case of bulk water where they converge to 1
for large $r$.
                                                                                                                            
Excess chemical potentials $\mu_{ex}$, for TIP3P water were evaluated
using Bennett's method of overlapping histograms \cite{B'76,HRN'01,
Vaitheeswaran_JCP:04}. The histograms of water insertion and removal (binding)
energies required for this calculation were collected in simulations with two
methanes in each droplet with no biasing potential applied to the methanes.

\section{Results and discussion}

We first characterize the equilibrium properties of water confined to a sphere
to discern the role of the boundary. The boundary should predominantly affect
the layer of water molecules close to the surface.

{\it Density profiles of confined water}: 
Water density profiles without methanes, which measure the local density
divided by the average density $\rho_0$, in the droplet ($\rho_0 =
N_w/V$ where $V$ is the total volume of the droplet) are shown in
Fig.~\ref{fig:dens_prof}. Remnants of layering, that are expected in the
presence of the walls, are apparent. Three layers can be seen for the droplets with $D=
2$, 3 and 4 nm that become progressively less well defined as the droplet size increases.
The hard walls confining the droplets induce a narrow region at the surface
in which the water density is substantially depleted - at the walls, the
density is $\approx 60\%$ of the average density, $\rho_0$. This results in
the scaled density being $> 1$ far from the walls (Fig.~\ref{fig:dens_prof}).
Confinement in a droplet perturbs the water structure and the perturbation
becomes progressively weaker as the droplet size increases.
In the $D=1$ nm droplet the local density is spatially different from the
average density $\rho_0$ at all values of $r$, the distance from the droplet
center.

Radial distribution functions for the confined water (Figs.~S.1 and S.2 in
Supporting information) are identical
to those for bulk TIP3P water \cite{JCMIK'83} in all important respects.
Confinement causes the $g(r)$'s to go to 0 at a distance $R+r_i$. In the case
of the 3 and 4 nm droplets the $g(r)$'s
are greater than 1 for $0.7\le r \le R-r_i$, the range in which there are
no oscillations in the functions and the normalization holds. This is due to
the fact that the scaled density (Fig.~\ref{fig:dens_prof}) is greater than 1.

Fig.~\ref{fig:BE}a shows the binding (or removal) energy distributions for all
water molecules in droplets of different sizes. The binding energy is the potential
energy required to remove a water molecule from the spherical droplet. As the
droplet size increases, the peak of the distribution shifts to larger negative values of $U$.
i.e. the waters inside become more strongly bound. The distribution for the
4 nm droplet is almost identical to that for bulk TIP3P water \cite{HRN'01}.
By this measure, the largest droplets in this study are substantially
bulk-like.

Fig.~\ref{fig:BE}b shows the same quantity evaluated for only the waters in
the first layer (defined to be 0.2 nm thick) closest to the droplet surface.
The distributions in Figs.~\ref{fig:BE}a and b are virtually identical for
the $D=1$ nm droplet. Thus, for small droplets there is essentially no
distinction between ``surface'' waters and water molecules in the interior.
For the $D=2$ nm droplet, the two distributions in Figs.~\ref{fig:BE}a and b
are similar but not identical. The distribution for the surface waters in
Fig.~\ref{fig:BE}b peaks at $\sim -65$ kJ/mol, while the corresponding one
evaluated for all the waters in the droplet (Fig.~\ref{fig:BE}a) peaks at
$\sim -70$ kJ/mol - a difference of $\sim 2 k_BT$, where $k_B$ is Boltzmann's
constant.

For larger droplets with $D=3$ and 4 nm, the surface waters
are substantially more loosely bound than those in the interior. We also note
that the distributions for the surface waters (Fig.~\ref{fig:BE}b) in the
3 and 4 nm droplets coincide with each other. i.e. even though the
4 nm droplet is more bulk-like than the 3 nm droplet (Fig.~\ref{fig:BE}a),
their surfaces are energetically identical.

Fig.~S.3 (Supporting information) shows the probability distributions of
$\cos \theta$ for surface and interior water molecules for droplets of different
sizes. Here, $\theta$ is the angle between the dipole moment of each water molecule
and the position vector for its oxygen atom (the origin being at the center of the
droplet). Orientations of the interior water molecules are almost isotropic.
However, surface waters are preferentially oriented with their dipole moments nearly
orthogonal to the droplet radius \cite{Pratt_Pohorille_Chem_Rev:02}.
This implies that a typical water molecule at the surface
has one hydrogen atom pointing away from the bulk, in the process sacrificing its
ability to donate one hydrogen bond. This conclusion agrees with the results of
earlier molecular dynamics (MD) simulations of Lee et~al. \cite{Lee_McCammon_Rossky_JCP:84}
who probed the behavior of water near a hydrophobic wall.

The $D$-dependent excess chemical potentials $\mu_{ex}$ of TIP3P water are calculated from
overlapping histograms of water insertion and removal energies
(see Fig.~S.4, Supporting information). As expected, $\mu_{ex}$
decreases monotonically as $D$ increases and approaches the value
for bulk TIP3P water (-25.3 kJ/mol) \cite{Vaitheeswaran_JCP:04}.
However, even at $D=4$ nm $\mu_{ex}$ differs significantly from the bulk value.
It is clear that in equilibrium with bulk water, none of these cavities
will have an average bulk density $\rho_B=997$ kg/m$^3$.
The presence of the depleted region near the
droplet surface implies that the average density inside will be less than $\rho_B$
when the droplet is in equilibrium with bulk water.

Introduction of methane molecules into the droplets results in some
layering in the water around them. Fig.~S.5 (Supporting information) shows
methane-oxygen radial distribution functions. i.e. probability
distributions of methane-oxygen distances $P(r,dr)$, divided by $r^2/R^3$,
where $r$ is the distance between them and $R=0.5D$ is the droplet radius.
Two water layers can be seen around the methanes.

{\it Solvation of methane}:
Fig.~\ref{fig:free_energy_D1-4}a shows the free energy of a single methane
molecule as a function of its distance from the center in droplets of different diameters.
The position dependent free energy is evaluated using a harmonic biasing potential
to localize the methane molecules at different values of $r$, the distance from the
droplet center. Each value for $r$ corresponds to a shell concentric with the droplet
boundary. The final unbiased probability distributions (evaluated by WHAM, as described earlier)
are scaled by $(r/R)^2$ before taking the logarithm and multiplying by $-k_BT$
to obtain the free energy.   Thus, these profiles do not include the $4\pi r^2$
contribution to the entropy.

In droplets of all sizes the surface is more favorable for the methane than the
interior by $\sim 10-15$ kJ/mol ($4-6$ $k_BT$ at 298 K) \cite{Matubayasi_Levy_JPC:96}.
This value is similar to the free energy difference of a methane at a water liquid-vapor
interface \cite{Henin_JCP:04,Ashbaugh_Pethica_Lang:03}. In the largest and the most
bulk-like droplet, the free energy difference of $\sim 10$ kJ/mol is close to
the hydration free energy of a methane molecule calculated from simulations
and experiment \cite{Henin_JCP:04}. This supports our view
of the droplet surface inducing a vapor-like state of low density with broken
hydrogen bonds \cite{Wallqvist_Levy_JPCb:01}. In the 2, 3 and 4 nm droplets, the free energy rises steeply
from the minimum at the surface and reaches its interior value within 0.4 nm
of the droplet surface and stays constant in the droplet interior. The
region with the steep gradient in the free energy coincides almost exactly with
the first layer of water in each droplet (Fig.~\ref{fig:dens_prof}).
In the smallest droplet ($D=1$ nm), there is no interior region; the free
energy changes continuously over the whole distance.

The preference of the methane for the droplet surface over the interior can be
understood in terms of the disruption in the water hydrogen bond network at the
surface. The presence of broken H-bonds at the surface implies that the
solvent has to pay a smaller entropic cost in solvating the methane if it is confined
to the surface. In other words, there is less entropy lost in reordering the hydrogen bonds
in the immediate vicinity of the non-polar solute. We attribute the larger free energy
differences in 2 and 3 nm droplets to the greater distortion in water structure
and greater surface curvature
in these droplets. Our interpretation that methane prefers being localized at the
surface is consistent with the observation that even in bulk water small hydrophobic
solutes occupy regions in which water has the largest number of unsatisfied hydrogen bonds
\cite{Matubayasi_Levy_JPC:96}.

From the temperature dependence of the free energy difference in the $D=3$ nm
droplet, we infer that at $T=328$ K the surface is more favorable by about
1.2 kJ/mol compared to $T=298$ K (see Fig.~\ref{fig:free_energy_D1-4}b). Estimates for
$T \Delta S$ (12.4 kJ/mol at $T=313$ K) and $\Delta U$ (-1.8 kJ/mol) show that the surface
free energy is entropically dominated. Both the entropy and the enthalpy are favorable at
the surface of the droplet for methanes.

{\it Potentials of mean force between methanes}:
Since the methane molecules are confined to the surface of the water droplets,
their mutual solvent mediated interaction will reflect the
disruption in the water structure due to the confining walls. Fig.~\ref
{fig:pmf} shows the potentials of mean force (PMFs) for two methane molecules
in droplets at various $D$ values. Compared with bulk water \cite{Shimizu_JCP:00},
these show an increased tendency for the methanes to associate. These profiles
show only a single minimum at contact.  Somewhat surprisingly, we find that
the secondary, solvent separated minimum that appears in the bulk profile is
completely absent, even when $D=4$ nm. The absence of the solvent separated
minimum in the $D=4$ nm droplet, which is entirely due to the presence of boundaries,
is intriguing because the properties of water are bulk-like in all crucial respects.
The solvent separated minimum in the bulk corresponds to configurations where
the methanes are separated by a single water molecule that is hydrogen bonded
to other waters. In the immediate vicinity of the confining surface, where
water-water hydrogen bonds are disrupted, such configurations are strongly
unfavorable. The preference for being pinned at the surface is consistent with a view that
a methane is likely to be localized in regions with maximum unsatisfied hydrogen
bonds. In confined water droplets this situation is readily realized at the boundary.

A striking feature of Fig.~\ref{fig:pmf} is that the calculated PMFs are
independent of the size of the droplet. This is because the methanes approach
each other along the surface which is energetically similar in all the droplets
(Fig.~\ref{fig:BE}b).

{\it Charged solutes}:
The interaction between charged solutes M$_{q+}$ and M$_{q-}$ is interesting because
of the competing preferences for surface (due to excluded volume) and the interior
(due to solvation of charges). Such a situation naturally arises in proteins in which
we expect that the charged backbone and polar and charged residues prefer to be solvated
in bulk water while
hydrophobic residues prefer the surface. It is likely that in confined systems the
relative tendency towards surface localization and solvation in the interior of the
droplet would depend on the magnitude of $q^+$. The presence of charges on the solutes will reduce
the tendency for surface solvation by making the interior of the droplet enthalpically more
favorable. The magnitude of $q^+$ will determine the extent to which ions will be
solvated. In Fig.~\ref{fig:free_energy_multi}
we plot the free energy of M$_{q+}$ and M$_{q-}$ with $q^+ = |q^-| = 0.4e$ as a
function of their radial position in droplets of different sizes.
The free energy profiles show a reduced preference for the surface for
ions of either sign compared to methane.

There is a dramatic dependence of the free
energy profiles of charged ions on $D$. This finding differs from the behavior of methane
which is always found at the surface (Fig.~\ref{fig:free_energy_multi}). For $D \le 3$ nm
the ions are more likely to be localized near the surface when $q^+ = |q^-| = 0.4e$ (see
Figs.~\ref{fig:free_energy_multi}a-c). For the largest droplet (Fig.~
\ref{fig:free_energy_multi}d), the interior which is almost completely bulk-like,
is more favorable for both positive and negative ions.  The penetration of the ions
into the droplet interior increases with decreasing curvature of the droplet surface, in accord
with the findings of Stuart et al. for the chloride ion \cite{Stuart_Berne_JPCa:99}.
Just as in bulk water \cite{L-Bell_Rasaiah:JCP:97} there is an asymmetry between the behavior of  cations and anions.
As $D$ increases, the enthalpy gain due to the solvation of the ions is greater than
the entropy loss due to ordering of water molecules around this solute. The enthalpy-entropy
interplay depends on $q^+$. In addition, the magnitude of the effect depends on
whether the ion is positively or negatively charged. For example, for $q^+ = 0.4e$ in $D=4$
(Fig.~\ref{fig:free_energy_multi}d), M$_{q-}$ is only marginally more stable in the interior
of the droplet compared to the surface. On the other hand, M$_{q+}$ is strongly solvated in
the interior.   Anions have a stronger tendency for surface solvation than
cations with the same charge magnitude. Localization of negative ions at
the surface enables interaction of water molecules with unsatisfied hydrogen bonds.

{\it Solute charge density determines the extent of interior solvation}: On general theoretical
grounds it can be argued that charge density ($\frac{q^+}{v_s}$ where $v_s$ is the volume of the
solute) determines hydration of ions.  In the case of confined water charge density should determine
the preference of the ions for the surface compared to the interior. In order to probe the effect
of charge density on ionic interactions we also examined M$_{q+/-}$ and M*$_{q+/-}$
with $q^+=e$ (Fig.~\ref{fig:D3_q1.0}) in a $D=3$ nm droplet. The latter have twice the volume and therefore
half the charge density of the former.  Both  M$_{q+}$
and M$_{q-}$ have a strong (enthalpic) preference for the interior
over the surface of the droplet, with the free energy difference being of the order
of $10-15$ $k_BT$. Molecular dynamics  simulations with a non-polarizable
force field \cite{Marrink_Lang:01} have obtained similar 
values for the  the free energies of sodium and
chloride ions in water near hydrophobic or purely repulsive surfaces and also
at a water liquid-vapor interface.
For M*$_{q+}$ and M*$_{q-}$, the surface is much less
unfavorable compared to M$_{q+}$ and M$_{q-}$ respectively. The profile for M*$_{q-}$ shows that
the anion preferentially resides ~0.2 nm below the droplet surface, but the barrier to penetrate
the droplet interior is only on  the order of $k_BT$. Clearly, the solute charge density 
is important, not just the magnitude of the charge. Whether a solute molecule is solvated
at the surface or in the interior depends on the balance between solute-solvent and solvent-solvent
interaction energies \cite{Jungwirth_Tobias_JPCb:02}. Decreasing the charge density increases the
tendency for surface solvation.

The asymmetry in the solvation of cations and anions in water droplets, along with the charge
density dependence can also explain the differences in the solvation of sodium and halide ions
in water clusters and slabs \cite{Jungwirth_Tobias_JPCb:02}. Sodium and fluoride ions were
found to be solvated in the interior while the larger halides had a propensity for surface
solvation in the order Cl$^- <$ Br$^- <$ I$^-$. The tendency for surface solvation in these
ions apparently correlates with their polarizabilities \cite{Jungwirth_Tobias_JPCb:02}, but also their
charge densities and the sign of the charge. The current results readily explain this trend
even without considering polarizability. From our results it is clear that decreasing charge
density increases the tendency for surface solvation. For a given charge density, anions have a
greater preference for the surface than cations.

Thus, the presence of electric charge on
a solute molecule drives it away from the
surface region where solvent hydrogen bonds are broken. Positive ions have a much reduced
preference for surface solvation compared to negative ions. This is also reflected in the
PMFs between   M$_{q+}$ and M$_{q-}$  charge magnitude 0.4$e$ which
are plotted in Fig.~\ref{fig:pmf_q}a. As the size of the water droplet increases,
the interior becomes increasingly bulk-like
and for the largest droplet, the profile is very similar to that for two methanes
in bulk water, with a solvent separated minimum at a separation of $\sim0.7$ nm.
In the 2 and 3 nm droplets, the secondary minimum is destabilized by $\sim 2$
kJ/mol relative to the 4 nm droplet, reflecting the greater role of confinement.

Fig.~\ref{fig:pmf_q}b shows the PMFs  between M$_{q+}$ and
M$_{q-}$ of charge magnitude $e$. In the 3 and 4 nm droplets, the profiles are virtually
identical. This is not surprising since these ions are strongly driven away from the surface
even in droplets of 3 nm diameter.
The secondary minimum is slightly less favorable (by $\sim$ 0.5 $k_B T$) in the 2 nm droplet
reflecting the effect of confinement.

 Fig.~\ref{fig:PMF_d3_q1.0} compares the PMFs between M$_{q+}$ and M$_{q-}$ and
M*$_{q+}$ and M*$_{q-}$ ($q^+=e$), in the 3 nm droplet. As M*$_{q+}$ and M*$_{q-}$ are driven closer
to the surface, as a result of the  the reduced charge density, the secondary, solvent separated minimum in the PMF
becomes shallower.

\section{Conclusions}
Our results show that methane molecules confined in spherical water droplets  have a strong preference for the surface, thereby maximizing the solvent entropy.
This effect is due to the extensive disruption of water hydrogen bonds
caused by the confining walls. The PMF between two methane molecules confined in these
droplets reflects the disruption in solvent structure caused by the confinement.
These profiles show only a single minimum at contact; the secondary,
solvent separated minimum seen in bulk water is completely absent. Confinement
by non-polar walls  increases the tendency of hydrophobic solutes to associate.
Similar behavior   will occur in the presence of non-polar surfaces immersed in bulk
water.  Since surfaces (confining or otherwise) are almost always present in practical situations,
we expect the PMF of Fig.~\ref{fig:pmf} to reflect methane-methane interactions in such cases. 
The addition of charges to the methanes reduces their preference for surface solvation.
Correspondingly, the PMF between M$_{q+}$ and M$_{q-}$ becomes progressively less
influenced by the bounding surfaces as the droplet size and the magnitude of charge increase. The balance between
interior solvation and preference for the surface is determined by the charge density of the solute.  As the solute charge
density decreases the boundary of the droplet becomes less unfavorable.  Our results show that the enhanced 
preference of the heavier halides for interfacial water can be fully rationalized in terms of decreased charge density.  

The primary motivation for computing the PMF between M$_{q+}$ M$_{q-}$ molecules is to understand
the effect of confinement when both charged (or polar) and hydrophobic interactions are
simultaneously present. Such is the case in proteins which are made up of hydrophobic
($\approx55\%$) residues and polar and charged residues ($\approx45\%$).
The results for M$_{q+}$ M$_{q-}$ suggests confinement effects on the stability of the
folded state depend on the subtle interplay between hydrophobic and charged
interactions. Let $\epsilon_H$ and $\epsilon_P$ denote the energies of confinement-induced
hydrophobic and polar interactions respectively. We predict that when $\epsilon_P/\epsilon_H>1$ then,
under folding conditions, the native state should be entropically stabilized
\cite{Betancourt_Thirumalai_JMB:99}. In the
opposite limit, $\epsilon_H/\epsilon_P>1$, we expect destabilization of the folded state
with respect to the bulk. When hydrophobic interactions dominate then it is likely that the
polypeptide chain is pinned at the surface which invariably destabilizes the folded structure.
These qualitative predictions are consistent with explicit
simulations of poly-alanine models confined in carbon nanotubes \cite{O'Brien_unpub}.
We find that in this case, which corresponds to $\epsilon_H/\epsilon_P>1$, the helix is
destabilized in the carbon nanotube.

The situation is complicated when $\epsilon_H/\epsilon_P \sim O(1)$
as appears to be the case when $q^+=0.4e$. Stability of the confined polypeptide chains
for which $\epsilon_H/\epsilon_P \sim O(1)$ may depend on sequence and the precise interaction
energetics between water, peptide and the walls \cite{Sorin_Pande_JACS:06,Cheung_Thirumalai_JMB:06}.
It should be emphasized that conformational
entropy of the polypeptide chain and the sequence will play a key role in determining the
stability of proteins under confinement. The present study suggests, in accord with the
conclusions reached elsewhere \cite{Cheung_Thirumalai_JMB:06}, that the diagram
of states for polypeptide chains is complicated especially when $\epsilon_H/\epsilon_P \sim O(1)$.

It has been proposed \cite{Rank_Baker_ProtSci:97,Cheung_Onuchic_PNAS:02} that barriers in the
folding process arise because water molecules that may be trapped between hydrophobic residues
have to be squeezed out prior to the formation of the hydrophobic core.
This conclusion is rationalized in terms of the PMF between methane
molecules in bulk water, in which there is a
barrier between the contact and solvent-separated minima.
Such an explanation is likely not applicable for folding in nanopores
because of the absence of the solvent-separated minimum between small hydrophobic species
in confined water. Our results imply that barriers to protein folding in nanopores should be
greatly reduced. If the native state is unaffected then the barrier reduction must arise
because of confinement induced changes in the denatured states.
The arguments given here show that alterations in the interactions between hydrophobic and
charged species in confined water can lead to many sequence-dependent
possibilities for folding
in nanopores.

\section{Acknowledgements}
S. V. thanks Ed O'Brien and Margaret Cheung for useful discussions. We are grateful to
John~E.~Straub for bringing Ref.~\citenum{Jungwirth_Tobias_JPCb:02} to our attention.
This work was supported in part by a grant from the National Science Foundation through CHE 05-14056.

\section{Supporting Information Available}
Water-water radial distribution functions, probability distributions of water
molecule orientations, water excess chemical potentials $\mu_{ex}$ and 
methane-oxygen radial distribution functions.
This material is available free of charge via the internet at 
http://pubs.acs.org.

\bibliography{Vaithee.master}

\newpage

\begin{table}[ht]
 \caption{Lennard-Jones parameters used in the simulations.
  The methane-oxygen parameters are obtained from the methane-methane
  \cite{Kalra_JPC:04} and oxygen-oxygen \cite{JCMIK'83} parameters
  by applying the Lorentz-Bertholet mixing rules.}
 \label{tab:LJ_parms}
\setlength{\tabcolsep}{5mm}
 \begin{tabular}{|c|c|c|}
   \hline\hline
   & $\epsilon$ [kJ/mol] & $\sigma$ [nm] \\ \hline
methane-methane & 1.23 ($\epsilon_{MM}$) & 0.373 ($\sigma_{MM}$)\\
oxygen-oxygen   & 0.64 ($\epsilon_{OO}$) & 0.315 ($\sigma_{OO}$)\\
methane-oxygen  & 0.887 ($\epsilon_{MO}$) & 0.344 ($\sigma_{MO}$)\\
   \hline\hline
 \end{tabular}
\end{table}

\newpage

\begin{table}[ht]
 \caption{Number of water molecules $N_w$ in droplets of different diameters $D$ and
  with varying number of solutes $N_m$.}
 \label{tab:N_w}
\setlength{\tabcolsep}{7mm}
 \begin{tabular}{|c|c|c|c|}
   \hline\hline
$D$ [nm] & $N_m=2$ & $N_m=1$ & $N_m=0$ \\\hline
1.0 & 6 & 12 & 17 \\
1.5 & 48 & - & - \\
2.0 & 128 & 134 & 140 \\
3.0 & 460 & 466 & 472 \\
4.0 & 1107 & 1112 & 1118 \\
   \hline\hline
 \end{tabular}
\end{table}

\newpage

\section{Figure Captions}

\begin{figure}[htbp]
   \centerline{\includegraphics[width=5in]{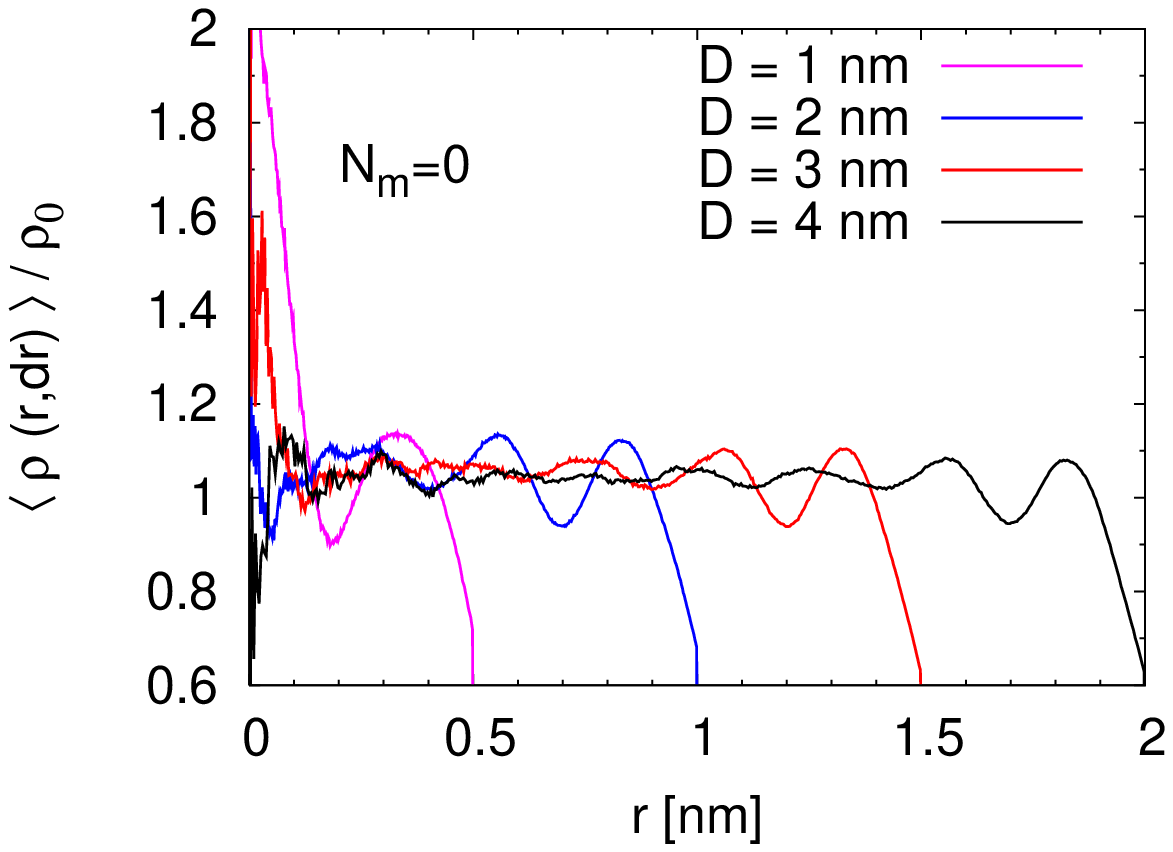}}
   \caption{}
   \label{fig:dens_prof}
 \end{figure}

Fig.~\ref{fig:dens_prof}: Density profiles of water in droplets of different sizes.
     There is depletion of water close to the surface of the sphere.
     $D$ is the droplet diameter in nm and $N_m$ is the number of
     solute molecules.

\begin{figure}[htbp]
   \centerline{\includegraphics[width=5in]{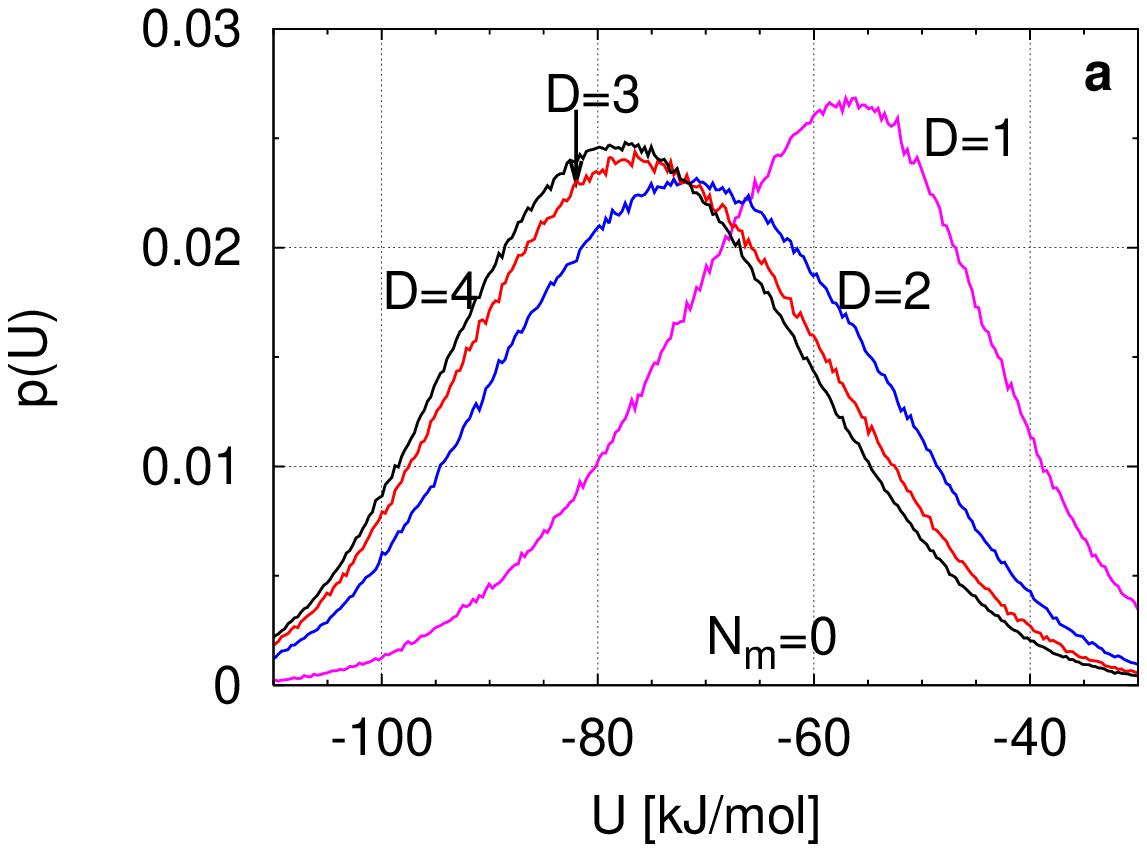}}
   \centerline{\includegraphics[width=5in]{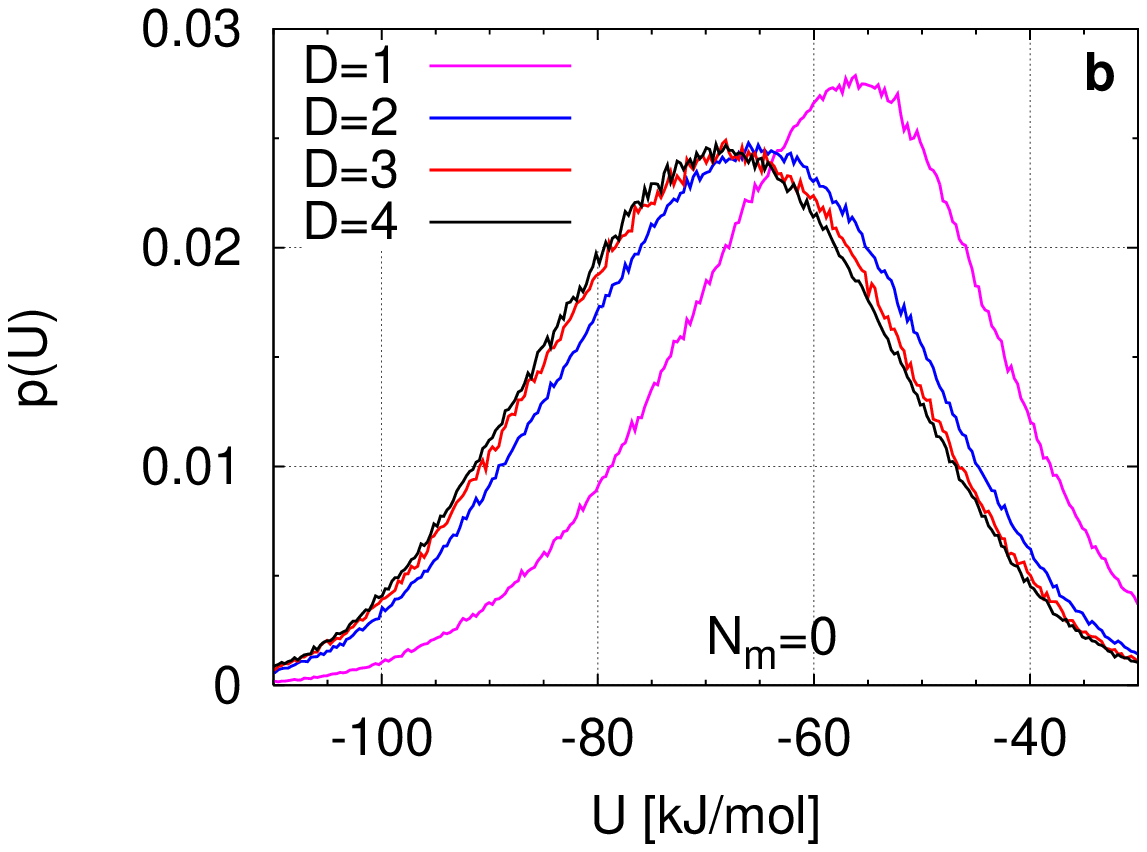}}
   \caption{}
   \label{fig:BE}
 \end{figure}

Fig.~\ref{fig:BE}: Probability distributions $p(U)$, of binding energies $U$ of water
    molecules. (a) The water molecules in the whole droplet are considered. (b) same
    as (a) except the distributions are computed using water molecules that are in the surface
    layer (oxygen atoms within 0.2 nm of droplet surface). Comparison of (a) and (b)
    shows that water molecules at the surface are loosely bound.

\begin{figure}[htbp]
   \centerline{\includegraphics[width=5in]{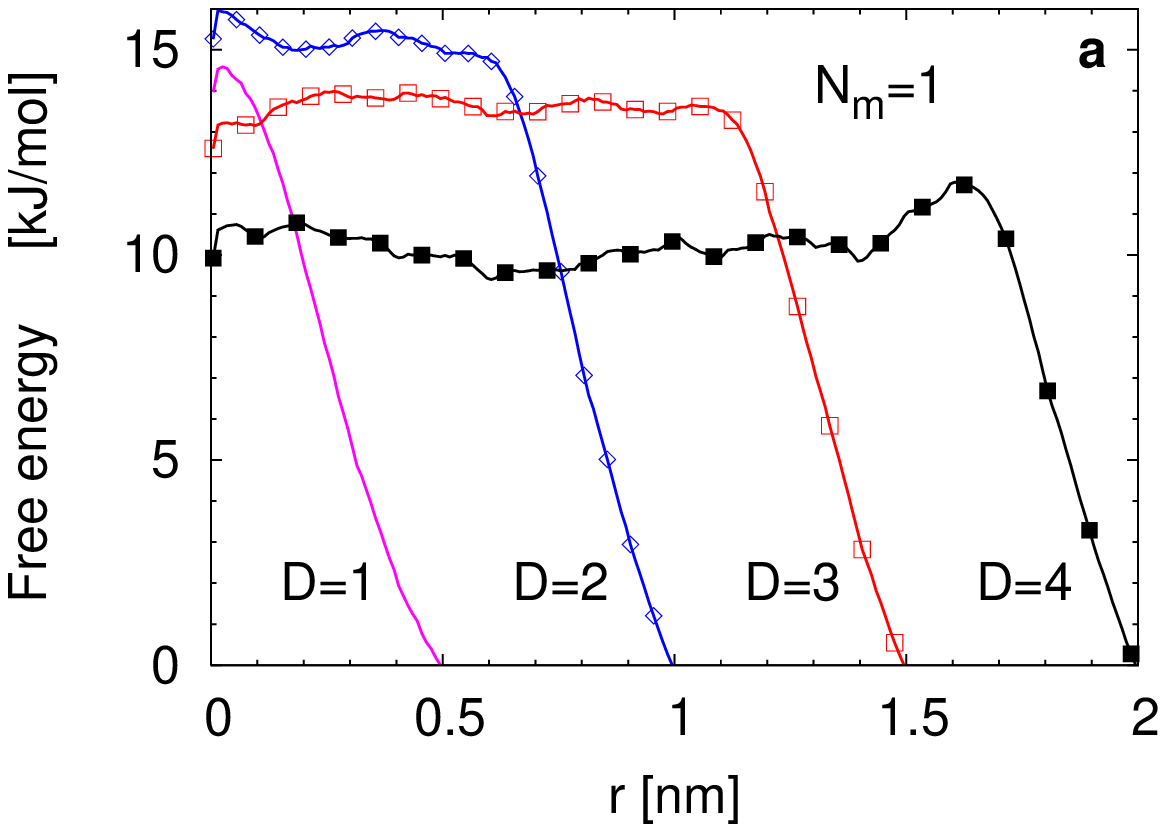}}
   \centerline{\includegraphics[width=5in]{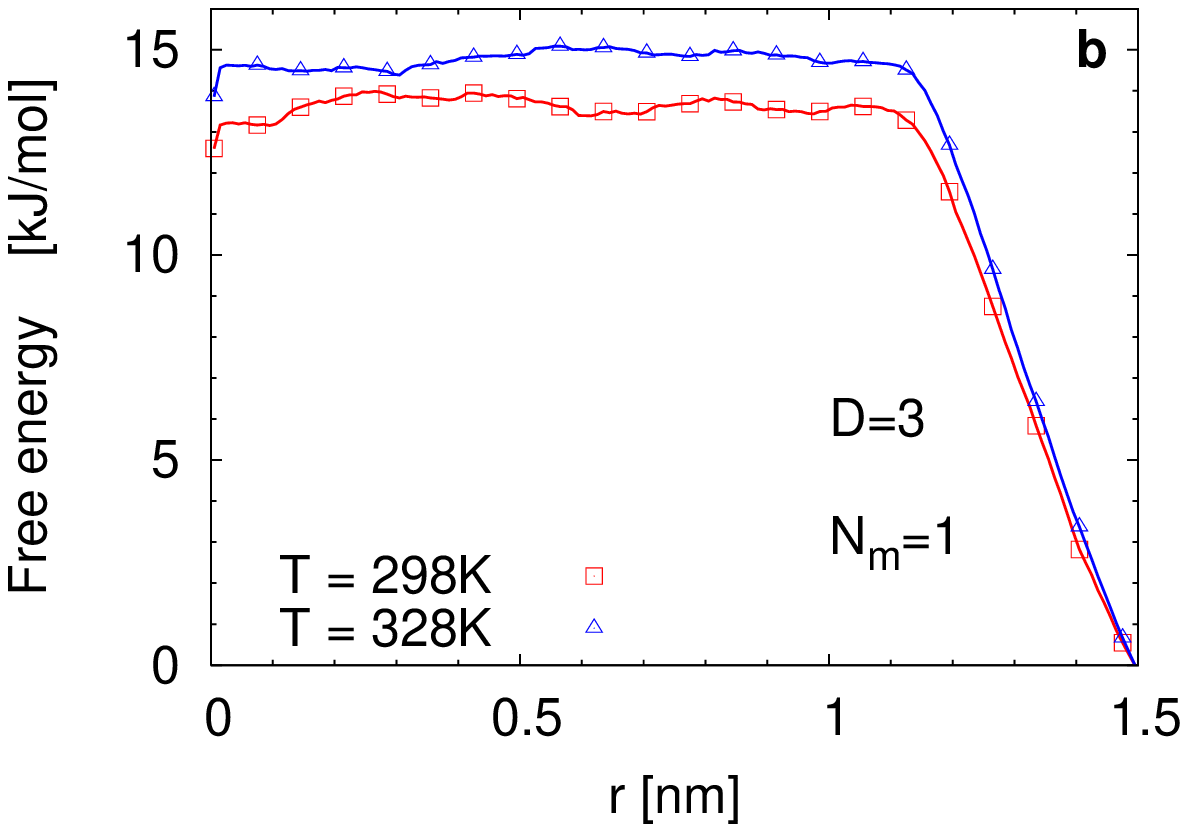}}
   \caption{}
   \label{fig:free_energy_D1-4}
 \end{figure}

Fig.~\ref{fig:free_energy_D1-4}: (a) Free energy of a single methane molecule as a function
     of its distance from the droplet center. The clear preference for the surface is
     evident at all values of $D$. (b) Temperature dependence of the free energy of a single methane
     molecule as a function of its distance from the center of a droplet of diameter 3
     nm. The zero of the free energy scale for every droplet is at its surface,
     and only differences within each curve are relevant.

\begin{figure}[htbp]
   \centerline{\includegraphics[width=5in]{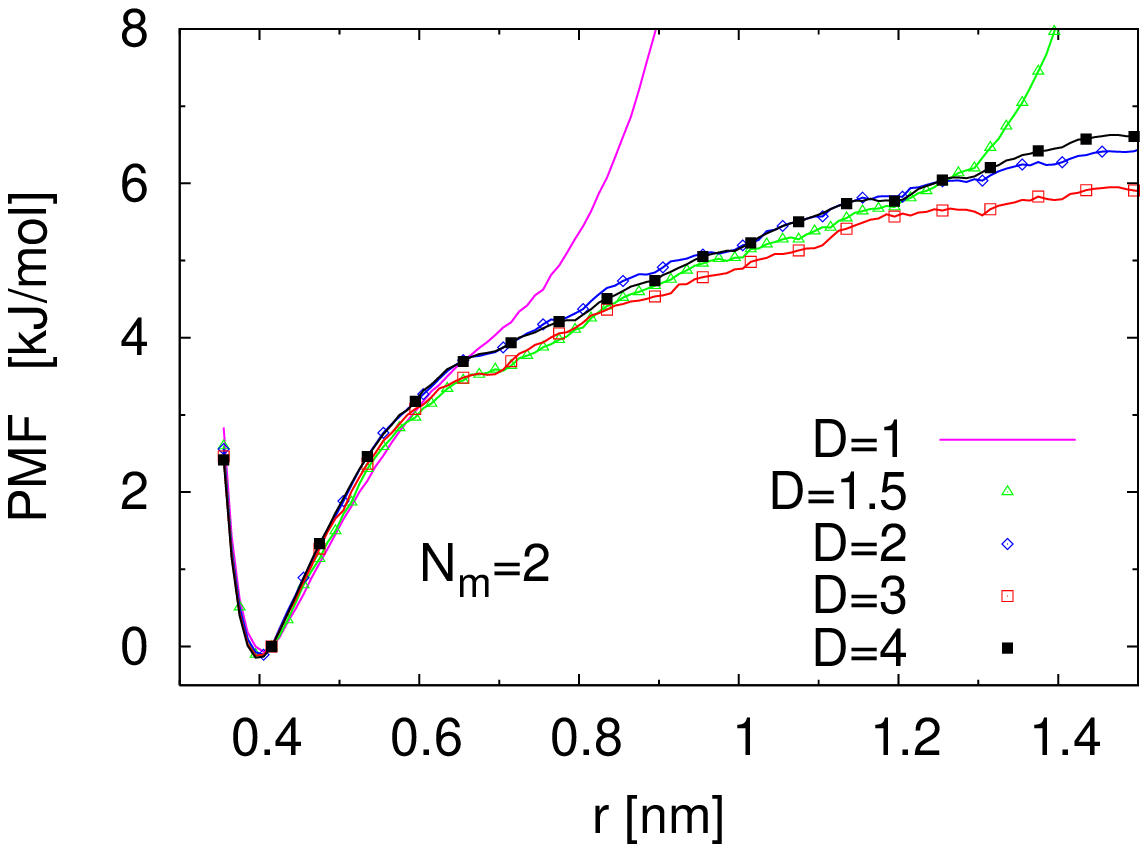}}
   \caption{}
   \label{fig:pmf}
 \end{figure}

Fig.~\ref{fig:pmf}: Potentials of mean force between two methane molecules
    in droplets of different sizes. Just as in the bulk, there is a distinct primary
    minimum. However, the characteristic solvent separated minimum is absent even at
    $D=4$ nm. The curves are shifted vertically so that the zero of the free energy
    scale is at contact for the two methanes. Only differences within each curve are relevant.

\begin{figure}[htbp]
   \centerline{\includegraphics[width=7in]{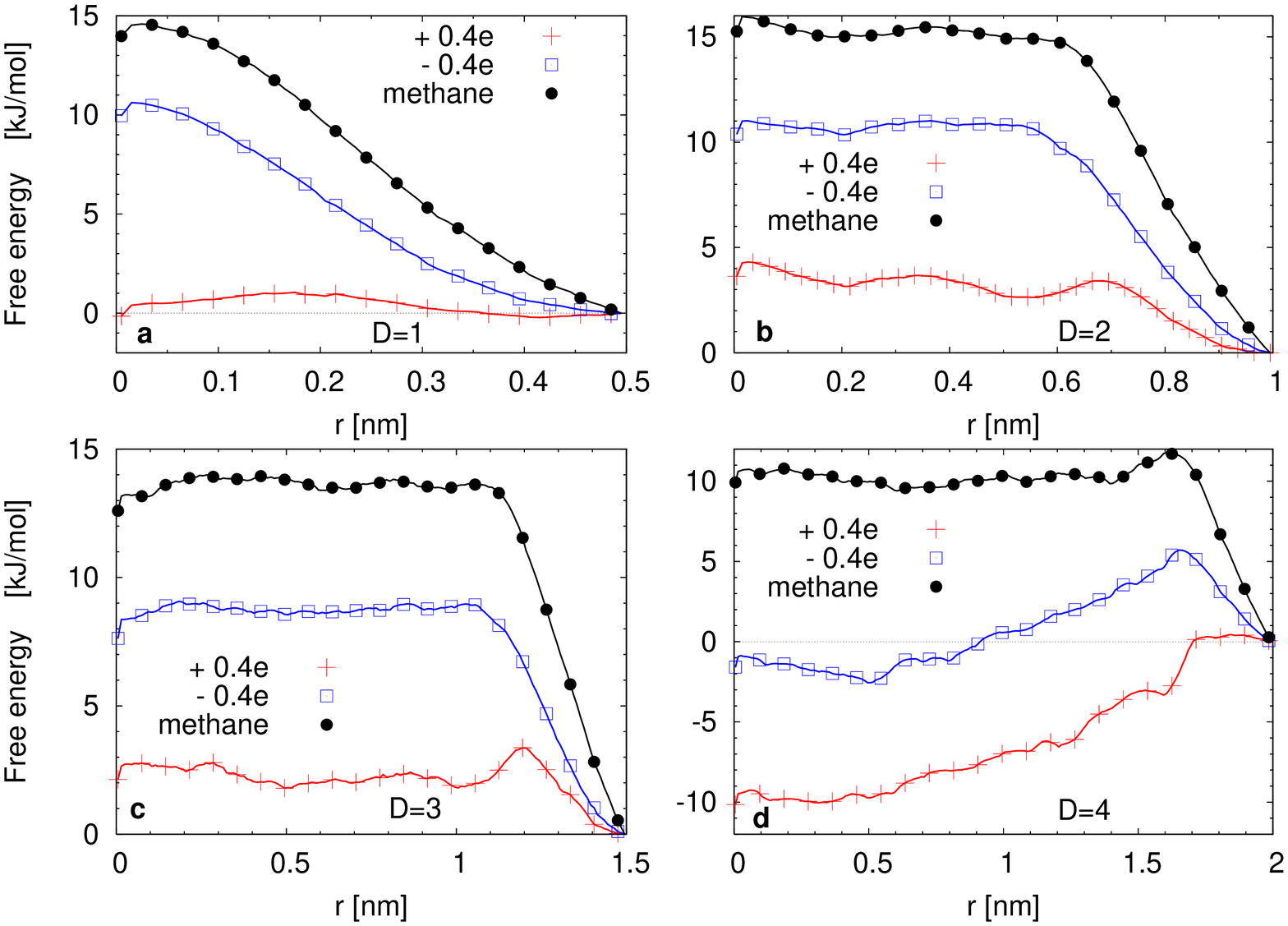}}
   \caption{}
   \label{fig:free_energy_multi}
 \end{figure}

Fig.~\ref{fig:free_energy_multi}: Free energies profiles for ions of charge magnitude
     0.4$e$ in droplets of diameter $D=1-4$ nm. Corresponding profiles for
     methane are included for comparison. At all values of $D$ there is an asymmetry,
     as in bulk water, between positive and negative charges. For 
     $q^+$ the preference for surface localization depends dramatically on the value of
     $D$ (compare (a), (b), (c) with results in (d)). Each curve is referenced to
     the surface of its respective droplet as in Fig.~\ref{fig:free_energy_D1-4}.

\begin{figure}[htbp]
   \centerline{\includegraphics[width=5in]{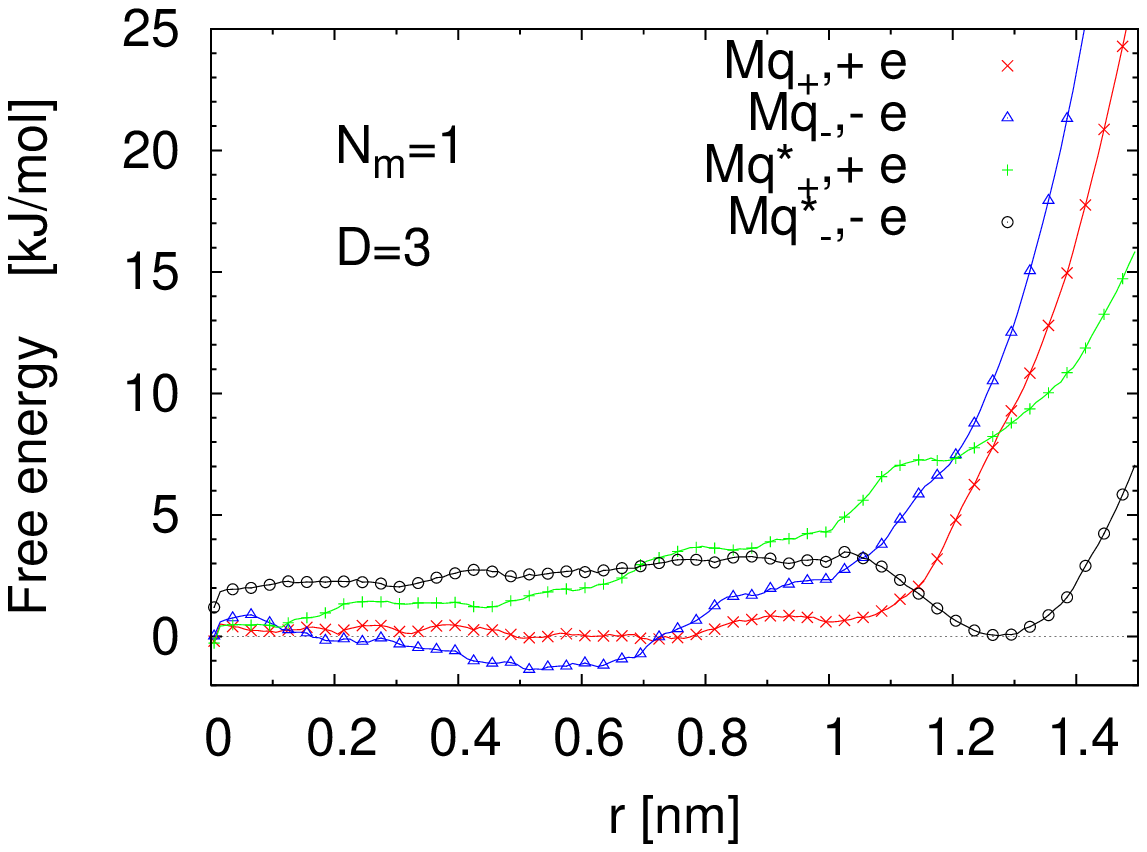}}
   \caption{}
   \label{fig:D3_q1.0}
 \end{figure}

Fig.~\ref{fig:D3_q1.0}: Free energies profiles for ions M$_{q+}$, M$_{q-}$, M*$_{q+}$
     and M*$_{q-}$ of charge magnitude $e$. The starred ions have twice the volume and
     therefore half the charge density of the corresponding unstarred ion. Curves for
     M$_{q+}$, M$_{q-}$ and M*$_{q+}$ are referenced to the origin (droplet center) while
     that for M*$_{q-}$ is referenced to the minimum at 1.3 nm for clarity.  As in
     Figs.~\ref{fig:free_energy_D1-4} and \ref{fig:free_energy_multi}, only differences
     within each curve are relevant.

\begin{figure}[htbp]
   \centerline{\includegraphics[width=5in]{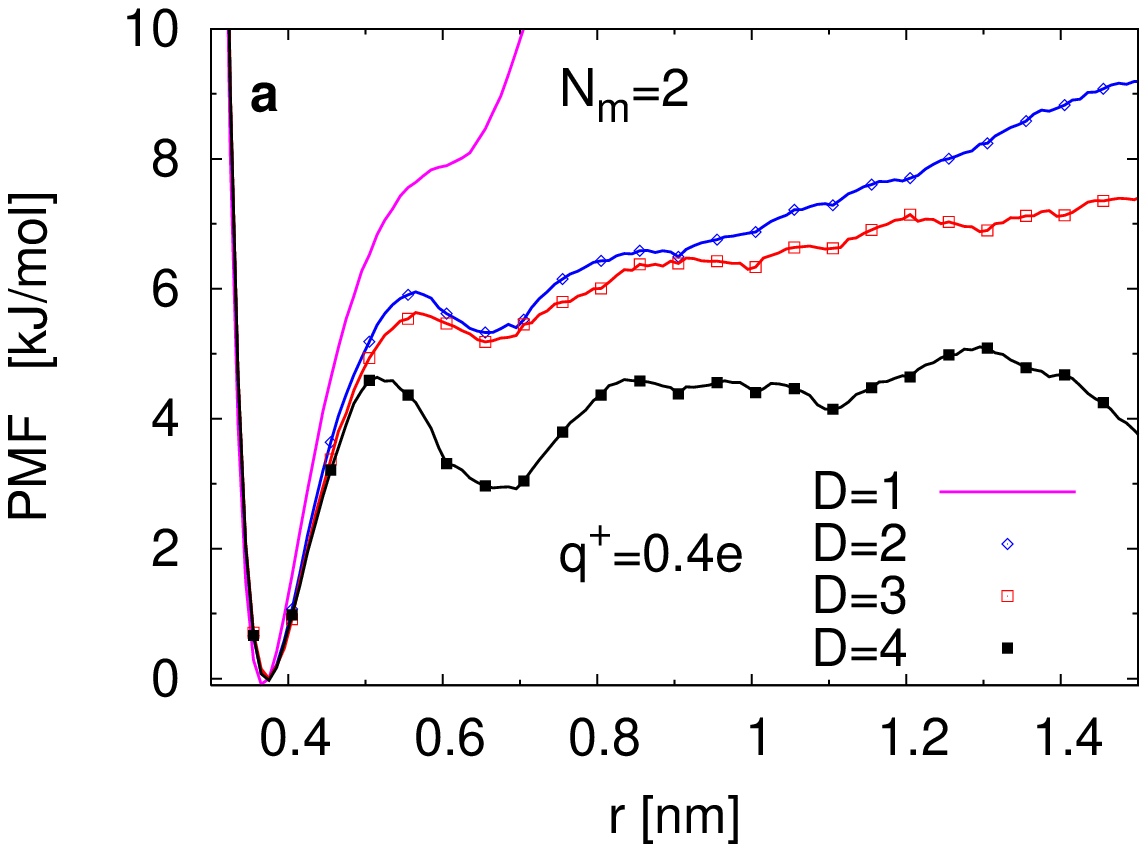}}
   \centerline{\includegraphics[width=5in]{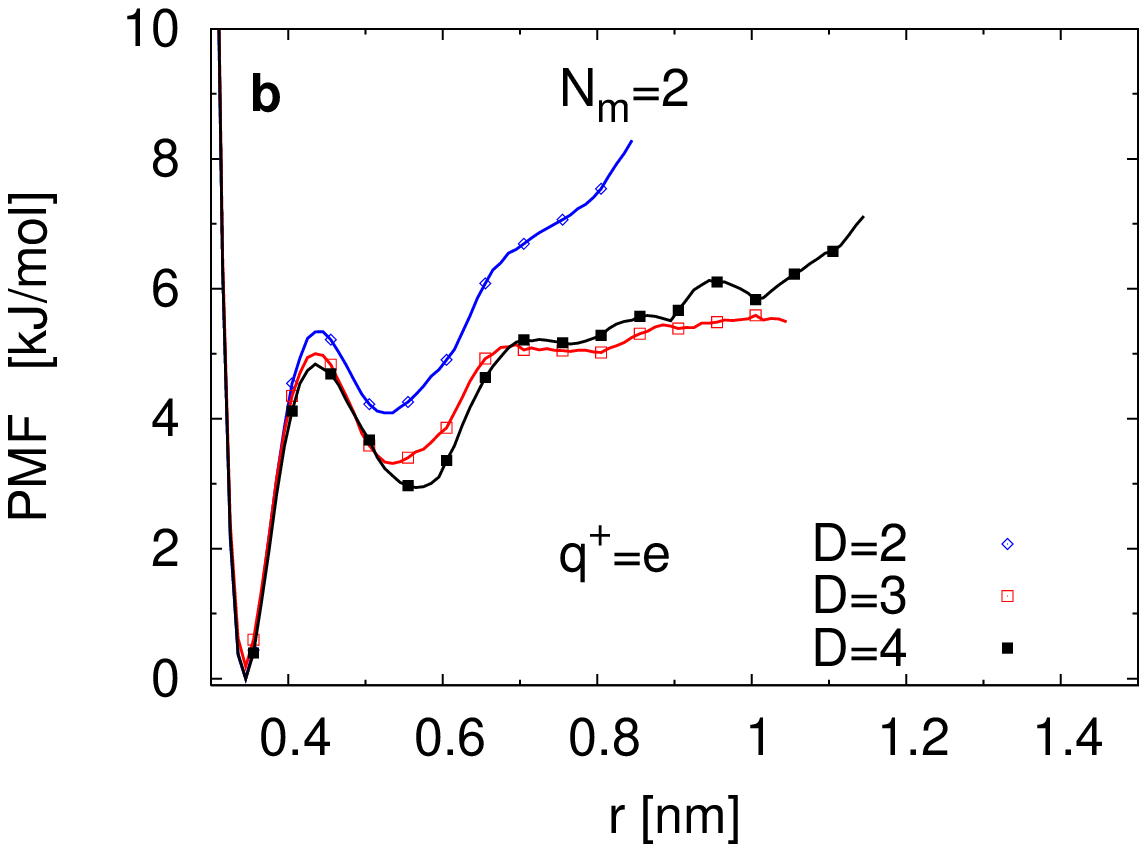}}
   \caption{}
   \label{fig:pmf_q}
 \end{figure}

Fig.~\ref{fig:pmf_q}: Potentials of mean force between M$_{q+}$ and M$_{q-}$. (a) The
     magnitude of the charge $q^+ = 0.4e$. The symbols correspond to different values
     of $D$. (b) Same as (a) except $q^+ = e$. In this case, due to the tendency of the
     charges to be fully solvated the PMF has bulk characteristics even at $D=3$ nm.
     Curves are shifted vertically as in Fig.~\ref{fig:pmf}.

\begin{figure}[htbp]
   \centerline{\includegraphics[width=5in]{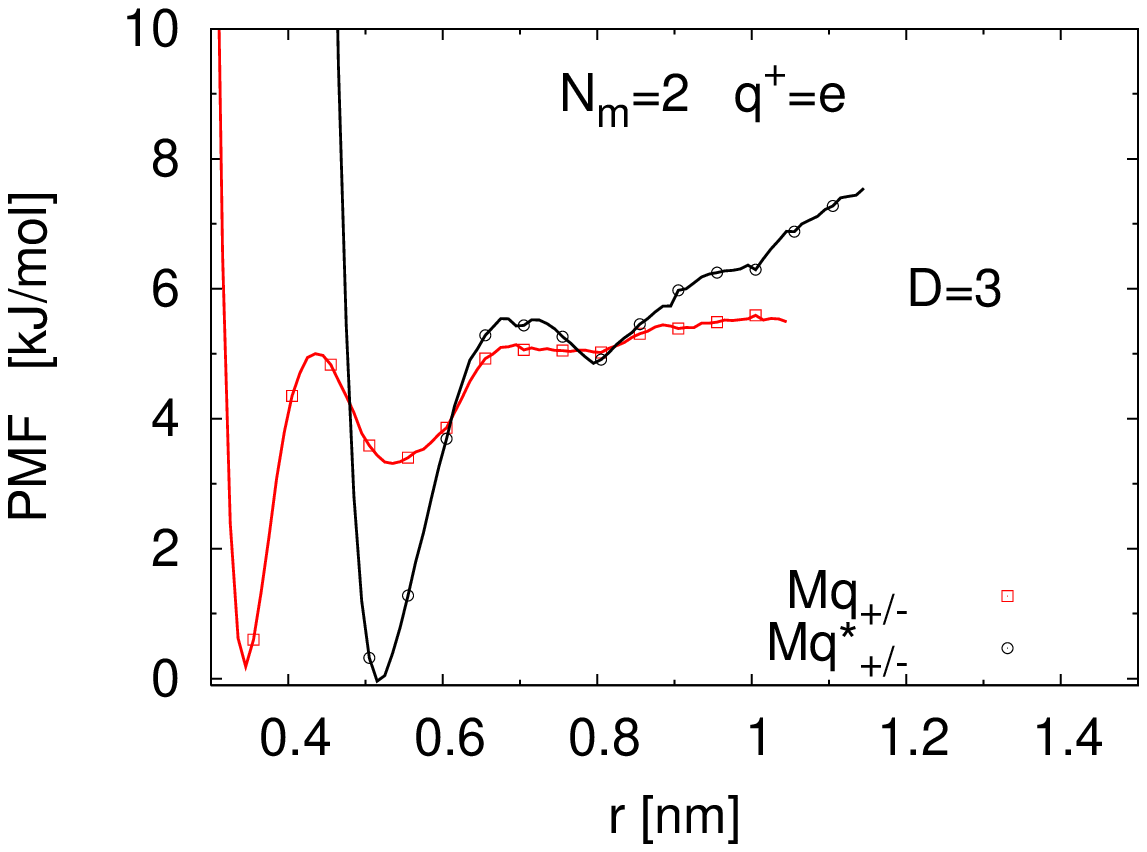}}
   \caption{}
   \label{fig:PMF_d3_q1.0}
 \end{figure}

Fig.~\ref{fig:PMF_d3_q1.0}: Potentials of mean force between M*$_{q+}$ and M*$_{q-}$
     ($q^+=e$) in the $D=3$ nm droplet (in black). The corresponding curve for M$_{q+}$ and M$_{q-}$
     is shown in red for comparison. Curves are shifted vertically as in Figs.~\ref{fig:pmf}
     and \ref{fig:pmf_q}.

\newpage

\renewcommand{\thefigure}{}

 \begin{figure}[htbp]
   \centerline{\includegraphics[]{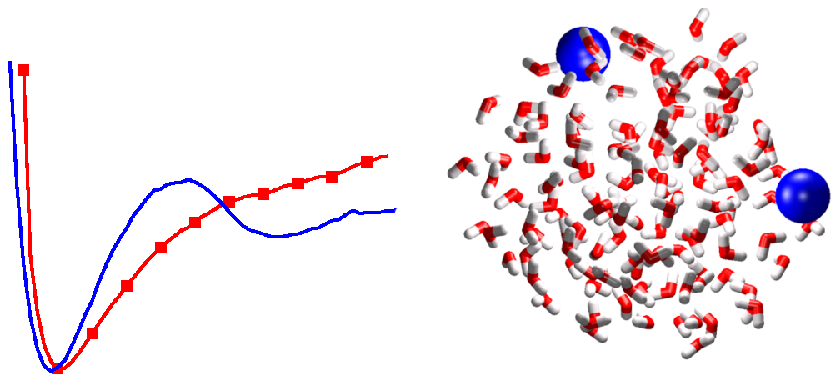}}
   \label{fig:TOC}
 \end{figure}

\end{document}


\fancyhead{}
\fancyfoot{}
\cfoot{S\thepage}

\renewcommand{\thefigure}{S.\arabic{figure}}

\title{Hydrophobic and Ionic Interactions in Nano-sized Water
Droplets \\- supporting information}
                                                                                                                             
\author{S. Vaitheeswaran$^{1}$ and D. Thirumalai$^{1,2,*}$,\\
$^1$Biophysics Program, Institute for Physical Science and Technology,\\
$^2$Department of Chemistry and Biochemistry,\\
University of Maryland, College Park, MD 20742\\
Email: {\tt thirum@glue.umd.edu}}
                                                                                                                             
\date{\today}

\maketitle
\thispagestyle{fancy}

  \begin{figure}[htbp]
    \centerline{\includegraphics[width=4in]{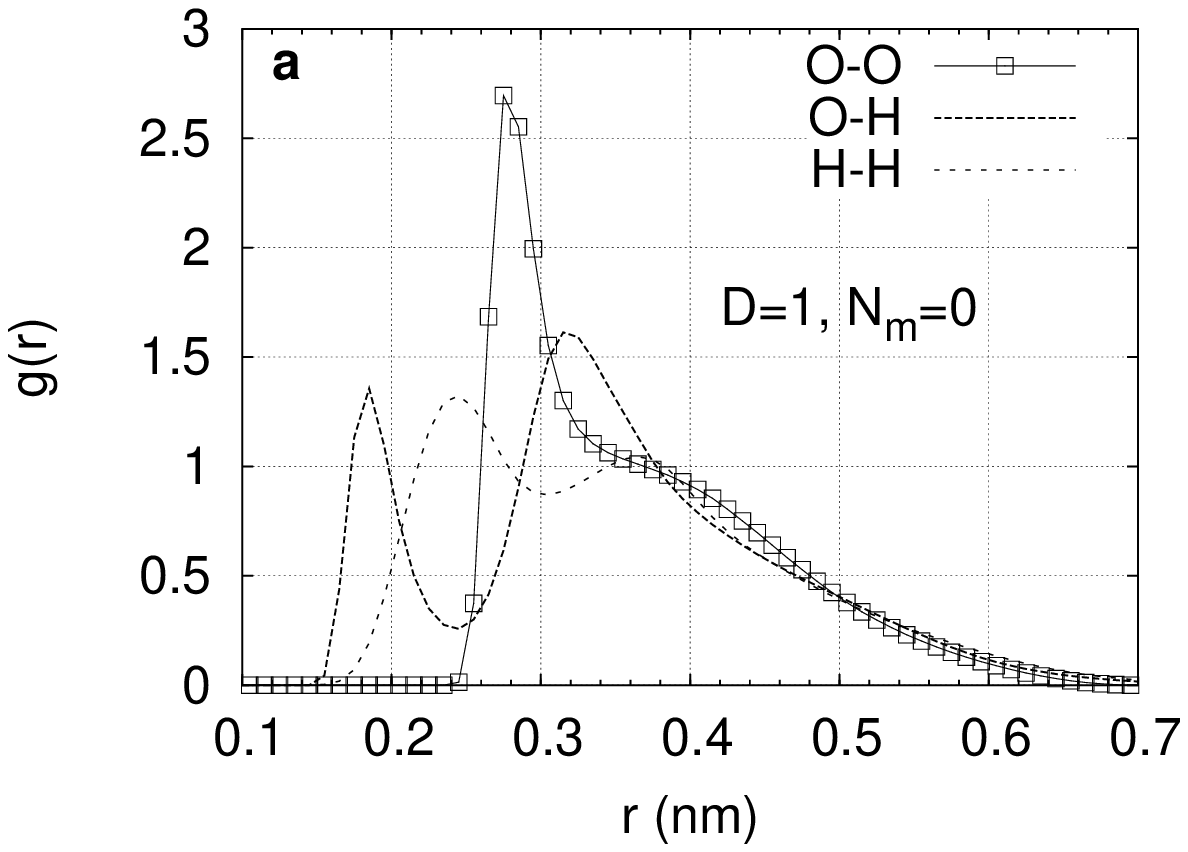}}
    \centerline{\includegraphics[width=4in]{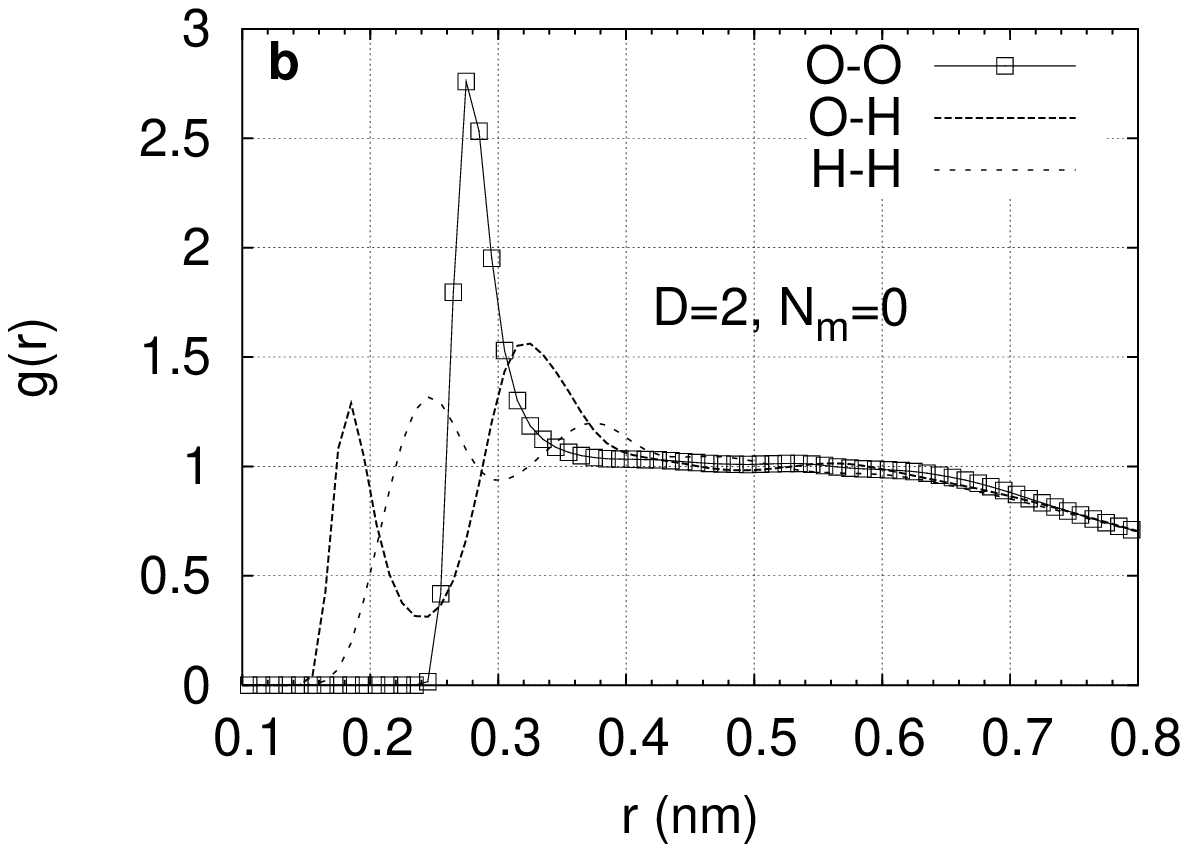}}
    \caption{Water-water radial distribution functions for droplets of
     diameter 1.0 and 2.0 nm. The normalization is accurate upto an
     oxygen-oxygen distance of 0.3 nm in (a) and 0.5 nm in (b).
     $D$ is the droplet diameter in nm and $N_m$ is the number of
     solute molecules.}
    \label{fig:gr_D1-2}
  \end{figure}

  \begin{figure}[htbp]
    \centerline{\includegraphics[width=4in]{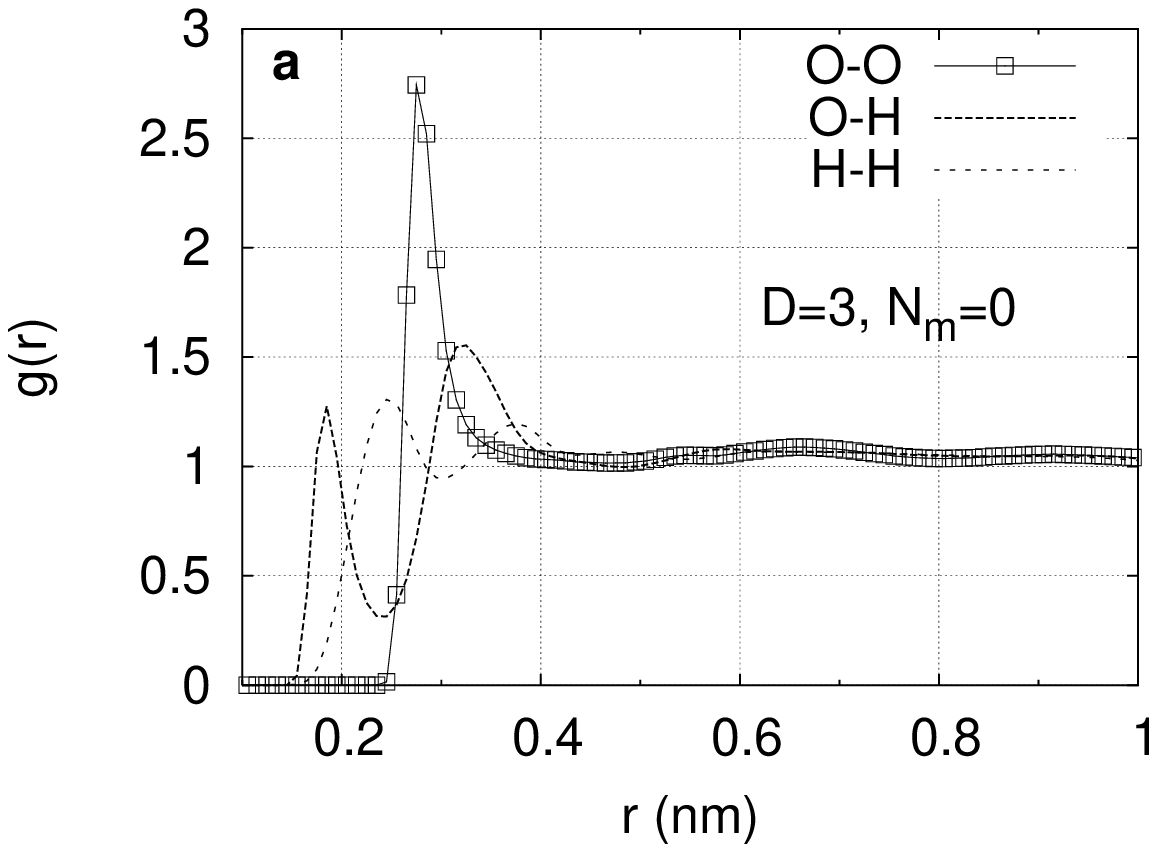}}
    \centerline{\includegraphics[width=4in]{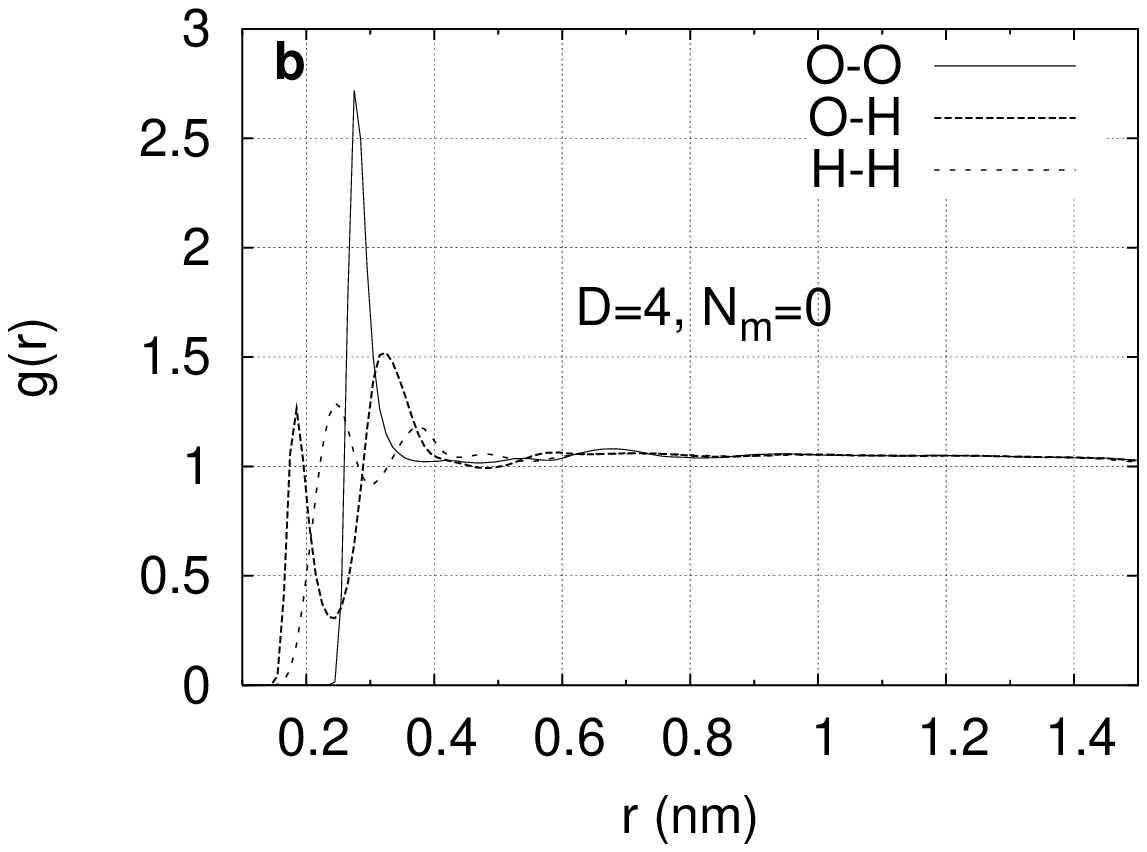}}
    \caption{Water-water radial distribution functions for droplets of
     diameter 3.0 and 4.0 nm. The normalization is accurate upto an
     oxygen-oxygen distance of 1 nm in (a) and 1.5 nm in (b).}
    \label{fig:gr_D3-4}
  \end{figure}

  \begin{figure}[htbp]
    \centerline{\includegraphics[width=5in]{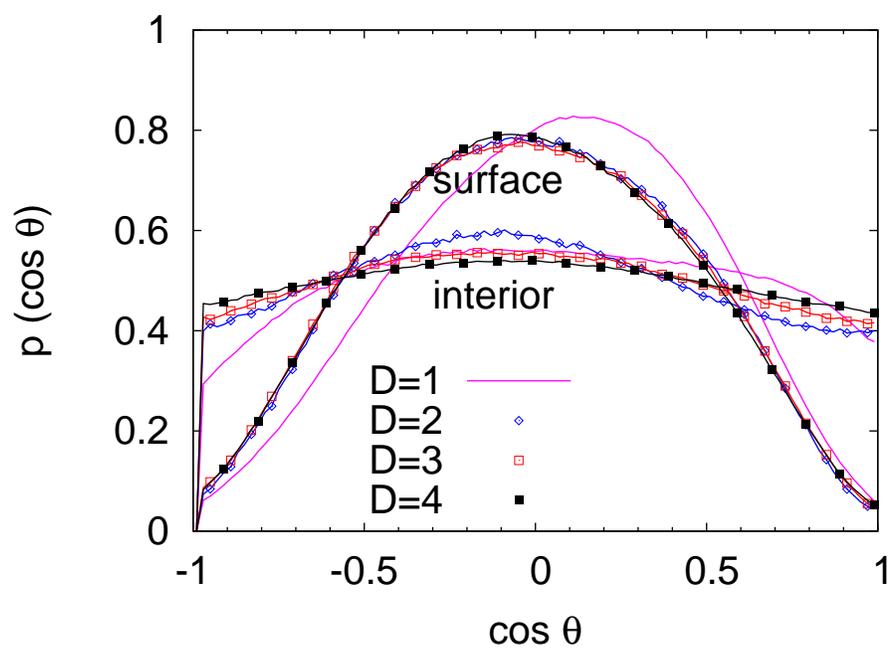}}
    \caption{Probability distributions of $\cos \theta$ for surface and
     interior waters for droplets of different sizes. $\theta$ is the angle
     between the dipole moment of each water molecule and the position vector
     of its oxygen atom.  The origin of the coordinate system is at the droplet
     center.}
    \label{fig:cos_theta}
  \end{figure}

  \begin{figure}[htbp]
    \centerline{\includegraphics[width=5in]{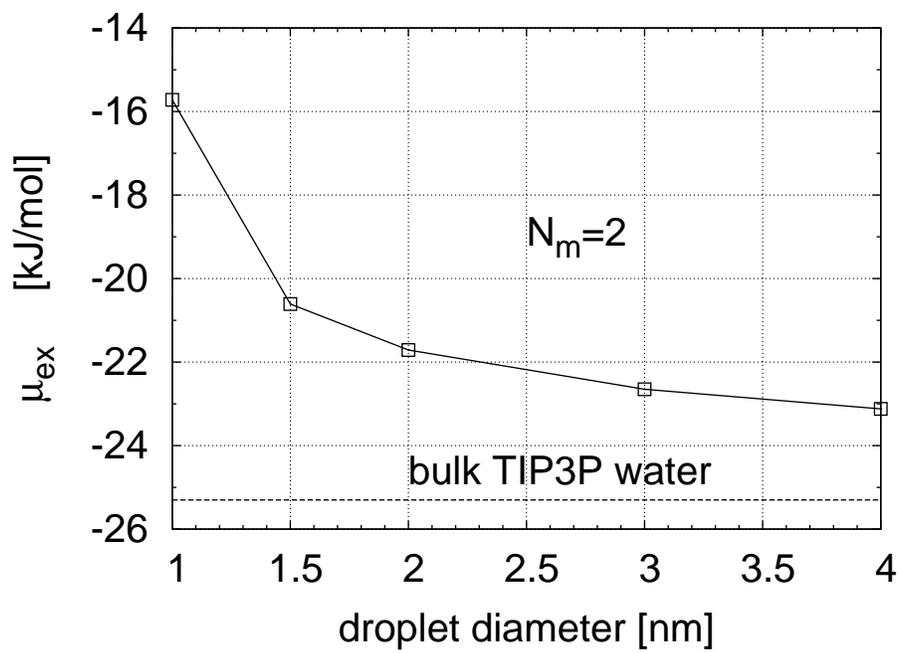}}
    \caption{Excess chemical potentials $\mu_{ex}$ of TIP3P water
     calculated by Bennett's method of overlapping histograms.}
    \label{fig:muex}
  \end{figure}

  \begin{figure}[htbp]
    \centerline{\includegraphics[width=5in]{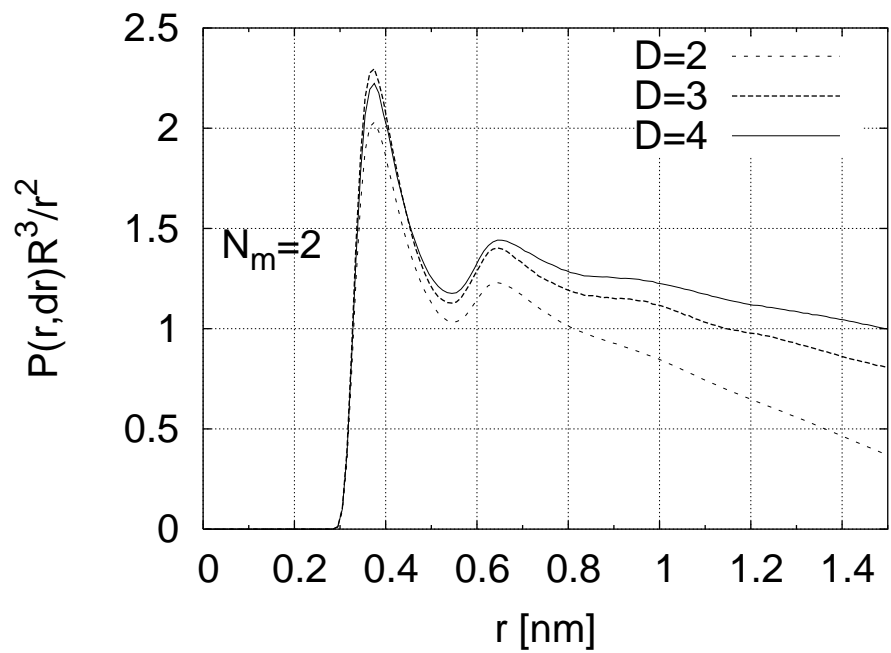}}
    \caption{Methane-oxygen radial distribution functions.
     $R=0.5D$ is the droplet radius. The $D=1$
     and $D=1.5$ nm droplets are omitted for clarity.}
    \label{fig:MO_dist}
  \end{figure}